\documentstyle[preprint,aps,psfig]{revtex}
\tighten
\begin{document}
\draft
\preprint{\vbox{Submitted to Physical Review C 
		          \hfill FSU-SCRI-99-07 \\}}
\title{Relativistic Treatment of Hypernuclear Decay} 
\author{L. Zhou and J. Piekarewicz}
\address{Department of Physics and 
         Supercomputer Computations Research Institute, \\
         Florida State University, Tallahassee, FL 32306}
\date{\today}
\maketitle
 
\begin{abstract}
We compute for the first time the decay width of $\Lambda$-hypernuclei
in a relativistic mean-field approximation to the Walecka model. Due
to the small mass difference between the $\Lambda$-hyperon and its
decay products---a nucleon and a pion---the mesonic component of the
decay is strongly Pauli blocked in the nuclear medium. Thus, the
in-medium decay becomes dominated by the non-mesonic, or two-body,
component of the decay. For this mode, the $\Lambda$-hyperon decays
into a nucleon and a spacelike nuclear excitation. In this work we
concentrate exclusively on the pion-like modes. By relying on the
analytic structure of the nucleon and pion propagators, we express the
non-mesonic component of the decay in terms of the spin-longitudinal
response function. This response has been constrained from precise
quasielastic $(\vec{p},\vec{n})$ measurements done at LAMPF. We
compute the spin-longitudinal response in a relativistic
random-phase-approximation model that reproduces accurately the
quasielastic data. By doing so, we obtain hypernuclear decay widths
that are considerably smaller---by factors of two or three---relative
to existing nonrelativistic calculations.
\end{abstract}
\vfill\eject

\narrowtext

\section{Introduction}
\label{sec:intro}
Understanding modifications to hadronic properties---such as masses 
and decay widths---in the nuclear medium occupies an important place 
in nuclear physics. The reason for such a status is the universal 
character of the problem. In electromagnetic phenomena one asks, 
for example, if the coupling of the photon to the 
proton gets modified in the nuclear environment. Similarly, hadronic 
processes---such as the $(\vec{p},\vec{n})$ reaction---search for 
evidence for new states of matter through the modifications of 
meson (e.g., pion) properties in the nuclear medium. Finally, 
the modification of masses and decay widths for vector mesons 
might provide, through relativistic heavy-ion experiments, 
conclusive evidence for the formation of the quark-gluon plasma.

In the present paper we are interested in understanding medium
modifications to the properties of the lambda-hyperon; in particular
modifications to its decay width. To our knowledge, this is the first
time that such a study will be carried out within the framework of a
relativistic mean-field approximation.  There are several reasons why
the $\Lambda$-decay is interesting.  First, the mesonic mode, $\Lambda
\rightarrow N\pi$, which comprises nearly 100\% of the decay in free
space, is unavailable in the medium because of Pauli
blocking~\cite{chepri53}. Second, a new nucleon-stimulated mode,
$\Lambda N \rightarrow NN$, opens up---precisely because of the
presence of the nuclear medium~\cite{dalitz58,blodal63}. As this
``non-mesonic'' mode dominates, the decay of the lambda in the nuclear
medium becomes sensitive to nuclear excitations. Finally, the decay is
interesting because the empirical $\Delta I\!=\!1/2$ rule---nearly 
exact for the free decay---appears to be violated in the nuclear
medium~\cite{oseram98}.

Because of simplicity, or perhaps because the weak $\Lambda N\pi$
couplings seem to be the only ones that are model 
independent~\cite{dubach86,dubach96}, most---although certainly 
not all---nonrelativistic calculations to date have concentrated 
exclusively on pion-like nuclear 
excitations~\cite{nieose93,osenie95,raossa95,alber95}. 
In this, our first, relativistic study of the decay we will 
proceed analogously; yet our results seem to indicate the need 
for the inclusion of additional nuclear 
excitations~\cite{dubach86,dubach96,ramben95}.

An important point that we will stress throughout this---and any 
future---publication is the need for consistency between
theoretical calculations and, seemingly unrelated, experimental 
data. For example, pion-like modes have been studied extensively 
by a variety of experiments. Arguably the most complete set of 
experiments were performed at the neutron-time-of-flight (NTOF)
facility at LAMPF~\cite{McCl92,Chen93}. These quasielastic 
$(\vec{p},\vec{n})$ measurements placed strong constraints on the 
(pion-like) spin-longitudinal response, by showing conclusively that 
the long-sought pion-condensed state does not exist. Although
improving fast, the present experimental situation on hypernuclear 
decay is still limited~\cite{grace85,szyman91,noumi95}. Thus, the 
theoretical guidance provided by alternative experiments should 
not be underestimated. Indeed, in this work we compute the decay 
of $\Lambda$-hypernuclei using the same exact model 
that was used successfully in describing the spin-longitudinal 
response extracted from the quasielastic $(\vec{p},\vec{n})$ 
experiments~\cite{horpie93}.

We have organized our paper as follows. In Sec.~\ref{sec:formal}
we develop the formalism needed to compute the in-medium decay
of the $\Lambda$-hyperon. We rely strongly on the analytic
structure of the nucleon and pion propagators to show that the
non-mesonic component of the decay is sensitive to the
spin-longitudinal response. The underlying $\Lambda N\pi$ dynamics 
is prescribed through an effective-Lagrangian approach. For the 
nucleon and lambda propagators we use a standard mean-field 
approximation to the Walecka model, while the pion propagator 
is evaluated in a random-phase approximation (RPA)---using a 
pseudovector representation for the $\pi NN$ vertex. Our results 
are presented in Sec.~\ref{sec:results}. Strong emphasis is 
placed on the dynamical quenching of the effective $\pi NN$ 
coupling in the nuclear medium, and on the corresponding 
reduction of the hypernuclear widths. Finally, in 
Sec.~\ref{sec:concl} we summarize our findings and make 
suggestions for future work.

\section{Formalism}
\label{sec:formal}
The dynamics of the lambda-nucleon-pion $(\Lambda N\pi)$ system is
described in terms of the following effective 
Lagrangian~\cite{dubach96}
\begin{equation}
 {\cal L}=g_{w}\bar{\psi}_{N}(x)\Big(1+\kappa\gamma^{5}\Big)
 {\boldmath\tau}\cdot{\boldmath\pi}(x){\psi}_{\Lambda}(x)+
 {\rm h.c.} \;,
 \label{Lagrangian}
\end{equation}
where $g_{w}\!=\!2.35\times10^{-7}$ is the weak $\Lambda N\pi$ 
coupling constant and $\kappa\!=\!-6.7$ is the ratio of the parity 
conserving to the parity violating coupling. Note that---in accordance 
to the empirical $\Delta I\!=\!1/2$  rule---the lambda-hyperon field 
occupies the lower entry of the two-component  ``isospinor'' 
${\psi}_{\Lambda}(x)$; the first entry is kept empty. This
effective Lagrangian will now be used to compute modifications to
the propagation of the $\Lambda$-hyperon in the nuclear medium.
Specifically, the modified $\Lambda$-propagator in symmetric
nuclear matter satisfies Dyson's equation~\cite{fetwal71}:
\begin{equation}
  S(p)=S_{0}(p)+S_{0}(p)\Sigma(p)S_{0}(p)+\ldots
      =S_{0}(p)+S_{0}(p)\Sigma(p)S(p)\;,
 \label{Lpropagator}
\end{equation}
where the lambda (proper) self-energy is defined by
\begin{equation}
 -i\Sigma(p)=3g_{w}^{2}
             \int \frac{d^{4}q}{(2\pi)^{4}}
             (1-\kappa\gamma^{5})G(p-q)
             (1+\kappa\gamma^{5})\Delta(q) \;.
 \label{Lselfenergy}
\end{equation}
Note that we have introduced the nucleon $(G)$ and pion 
$(\Delta)$ propagators. Moreover, the factor of ``3'' 
in front of the integral appears as a direct consequence 
of the $\Delta I\!=\!1/2$ rule---the $p\pi^{-}$ channel 
contributes twice as much as the $n\pi^{0}$ one. 

\subsection{Nucleon Propagator}
\label{sec:nucprop}
The nucleon propagator will be computed in nuclear matter using 
a mean-field approximation to the Walecka model~\cite{walecka74}.
Since most of the formal aspects of the derivation can be found 
in the textbook by Serot and Walecka~\cite{serwal86}---as well 
as in many other publications---we limit ourselves to a brief 
review of the central issues.

In a self-consistent (or Hartree) approximation the relativistic
nucleon propagator in nuclear matter may be written as
\begin{equation}
  G(k) = (\rlap\slash{k}^{\star}\!+\!M_{N}^{\star})
          \left[
           \frac{1}{k^{\star 2}\!-\!M_{N}^{\star 2}+i\eta}
          +\frac{i\pi}{E_{N}^{\star}({\bf k})}
           \theta(k_{F}-|{\bf k}|)
           \delta(k^{0}\!-\!E_{N}^{(+)}({\bf k}))  
          \right] \;, 
 \label{Npropagator}
\end{equation}
where the following quantities have been introduced:
\begin{equation}
   k^{\mu\star} \equiv(k^{0}\!-\!V_{N},{\bf k})   \;, \quad
   E_{N}^{(\pm)}({\bf k})\equiv E_{N}^{\star}({\bf k})\pm V_{N} \;, \quad
   E_{N}^{\star}({\bf k})\equiv \sqrt{{\bf k}^{2}+M_{N}^{\star 2}} \;.
 \label{Ndispersion}
\end{equation}
Note that once the Fermi momentum $k_{F}$ is specified, the effective
nucleon mass $M_{N}^{\star}$, the energy shift $V_{N}$---and with them
the full nucleon propagator---can be determined~\cite{serwal86}. The
three independent parameters of the Walecka model are the mass of the 
scalar meson, and the scalar and vector couplings respectively (the 
vector mass is kept fixed at its physical value of 783 MeV). These
three parameters---which are used to describe ground-state properties
of closed-shell nuclei throughout the periodic table---are 
given by:
\begin{equation}
      m_{\rm s}=520~{\rm MeV}   \;; \quad
      g_{\rm s}^{2}/4\pi=8.724  \;; \quad
      g_{\rm v}^{2}/4\pi=15.154 \;.
 \label{Walparam}
\end{equation}
The above nucleon propagator can be written alternatively in a 
spectral representation. That is, $G(k)\equiv G^{(+)}(k)+G^{(-)}(k)$, 
where
\begin{mathletters}
\begin{eqnarray}
  G^{(+)}(k) &=& 
          \sum_{s}\left(
           \frac{\theta(k_{F}-|{\bf k}|)}
                {k^0-E_{N}^{(+)}({\bf k})-i\eta}  +
           \frac{\theta(|{\bf k}|-k_{F})}
           {k^0-E_{N}^{(+)}({\bf k})+i\eta}\right)
           U^{\star}({\bf k},s)
           \overline{U}^{\star}({\bf k},s) \;, 
           \label{Npropagatorp} \\
  G^{(-)}(k) &=& 
          \sum_{s}
           \frac{V^{\star}({-\bf k},s)
                \overline{V}^{\star}({-\bf k},s)}
                {k^0+E_{N}^{(-)}({\bf k})-i\eta} \;. 
           \label{Npropagatorm}
\end{eqnarray}
\end{mathletters}
We have chosen to display the analytic structure of the nucleon
propagator to highlight the relevant physics that has been included 
at the mean-field level. For example, the $G^{(-)}$ term represents
the anti-particle (or filled Dirac sea) contribution to the nucleon 
propagator. This term is analytic in the full complex $k^{0}$-plane, 
except for the presence of negative-energy poles located (slightly) 
above the real axis. This component of the propagator does not 
contribute to the width of the $\Lambda$-hyperon in the nuclear 
medium. The $G^{(+)}$ term, in contrast, represents the
positive-energy part of the propagator. In addition to a change in 
the dispersion relation because of the presence of the mean fields,
the only other modification to this part of the propagator, relative 
to its free-space structure, is the shift in the position of the pole 
of all the states below the Fermi level from slightly below to
slightly above the real $k^{0}$-axis (see Fig.\ref{fig1}). In  this 
way, the ``quasinucleon'' propagator takes a form analogous to the 
one in free space---the conventional Feynman propagator---while at
the same time enforcing the Pauli-exclusion principle.

\subsection{Pion Propagator I.}
\label{sec:pionpropI}
Isolating the analytic structure of the pion propagator in nuclear
matter is not as straightforward as in the case of the above nucleon
propagator. Thus, in this case it is convenient to express the pion
propagator in its more general form---by using its Lehmann
representation~\cite{fetwal71}:
\begin{equation}
  \Delta(q) = \frac{1}{\pi}\int_{0}^{\infty}d\omega\,
              \frac{{\cal I}_{m}\Delta({\bf q},\omega)}
                   {\omega-q^{0}-i\eta}
  	    - \frac{1}{\pi}\int_{-\infty}^{0}d\omega\,
              \frac{{\cal I}_{m}\Delta({\bf q},\omega)}
                   {\omega-q^{0}+i\eta}
 \label{Lehmann}
\end{equation}
In this form, knowledge of the spectral weight (i.e., the imaginary
part of the propagator) is sufficient to reconstruct the full
propagator, albeit at the expense of introducing an additional
integral. Yet the virtue of such a representation is that the physical 
content of the imaginary part is simple and illuminating. We will
return to this point later in the section. Now we proceed to compute 
directly the imaginary part of the lambda self-energy. 

\subsection{Lambda Self-Energy}
\label{sec:LSelfEnergy}
As the lambda propagates through the nuclear medium it feels the
presence of the strong scalar and vector mean fields. Thus, the 
lambda propagator gets modified at the mean-field level in a 
manner similar to that of the nucleon propagator. However, 
because there are no pre-existing lambdas in the nucleus, there 
is no Pauli correction to the propagator. Hence, the lambda 
propagator looks exactly as a conventional Feynman propagator
\begin{equation} 
  S(p)=\frac{\rlap\slash{p}^{\star}\!+\!M_{\Lambda}^{\star}}
            {p^{\star 2}\!-\!M_{\Lambda}^{\star 2}+i\eta} \;, 
 \label{MFLPropagator}
\end{equation}
but with a mass and a dispersion relation modified by the strong
mean fields:
\begin{equation}
   p^{\mu\star} \equiv(p^{0}\!-\!V_{\Lambda},{\bf p})       
   \;, \quad
   E_{\Lambda}^{(\pm)}({\bf p})\equiv 
   E_{\Lambda}^{\star}({\bf p})\pm V_{\Lambda} 
   \;, \quad
   E_{\Lambda}^{\star}({\bf p})\equiv 
   \sqrt{{\bf p}^{2}+M_{\Lambda}^{\star 2}} \;.
 \label{Ldispersion}
\end{equation}
The only two parameters that remain to be specified are the 
coupling constants of the lambda to the scalar and vector 
fields. Here we determine them from the assumption that the 
strange quark in the lambda does not couple to the sigma nor 
to the omega meson. In this way we end up with the simple 
quark-model estimates of:
\begin{equation}
   g_{\sigma\Lambda\Lambda}/{g_{\sigma NN}}=
   g_{\omega\Lambda\Lambda}/{g_{\omega NN}}=2/3\;.
 \label{ratios}
\end{equation}
Of course other possible values---perhaps phenomenologically
more robust---can be easily incorporated in our calculation. 
Yet, irrespective of these values the effective lambda parameters 
become: 
\begin{equation}
    (M_{\Lambda}^{\star}-M_{\Lambda})=
    \left(\frac{g_{\sigma\Lambda\Lambda}}{g_{\sigma NN}}\right)
    (M_{N}^{\star}-M_{N}) \;; \quad
    V_{\Lambda}=
    \left(\frac{g_{\omega\Lambda\Lambda}}{g_{\omega NN}}\right)
    V_{N} \;.
 \label{Larams}
\end{equation}
In a mean-field approximation it is the strong scalar field 
that is responsible for inducing a  shift in the mass of the 
lambda, relative to its free-space value. Yet, its entire decay 
width is generated from the weak matrix element given in 
Eq.(\ref{Lselfenergy}); note that the dispersive contribution 
to the mass term is insignificant. Moreover, the contribution 
to the width comes entirely from the particle ($+i\eta$) 
component of the nucleon propagator [the last term in 
Eq.(\ref{Npropagatorp})]. This in turn constrains the 
contribution from the pion propagator to arise exclusively 
from the first ($-i\eta$) term in Eq.(\ref{Lehmann}); otherwise 
the $q^{0}$ integral in Eq.(\ref{Lselfenergy}) would vanish. 
In this way, the $q^{0}$ integral can be computed with the aid 
of Cauchy's theorem. We obtain,
\begin{eqnarray}
 {\cal I}_{m}\Sigma(p)
        =3g_{w}^{2}&&\int\!\frac{d^{3}q}{(2\pi)^{3}}
         \frac{(1\!-\!\kappa\gamma^{5})
        (\rlap\slash{k}^{\star}\!+\!M_{N}^{\star})
        (1\!+\!\kappa\gamma^{5})}{E_{N}^{\star}({\bf k})}
        \theta\Big(E_{N}^{\star}({\bf k})-E_{F}^{\star}\Big)
        \nonumber \\
	&&\int_{0}^{\infty}\!d\omega\,
	{\cal I}_{m}\,\Delta({\bf q},\omega)\,
        \delta\left(E_{\Lambda}^{(+)}({\bf p})\!-\!
        E_{N}^{(+)}({\bf k})\!-\!\omega\right) \;,
 \label{ImSigma}
\end{eqnarray}
where now $k^{\mu\star}$ is evaluated at its on-shell value:
$k^{\mu\star}\!=\!
(E_{N}^{\star}({\bf k}),{\bf k}\!\equiv\!{\bf p}\!-\!{\bf q})$.
To proceed, and in particular to isolate the Lorentz structure 
of the self-energy, it is convenient to use the following identity 
between gamma matrices:
\begin{equation}
  (1\!-\!\kappa\gamma^{5})
  (\rlap\slash{k}^{\star}\!+\!M_{N}^{\star})
  (1\!+\!\kappa\gamma^{5}) =
  (1\!-\!\kappa^{2})M_{N}^{\star}+
  (1\!+\!\kappa^{2})\rlap\slash{k}^{\star}+
  2\kappa\rlap\slash{k}^{\star}\gamma^{5} \;.
 \label{Ident}
\end{equation}
Note that a parity-violating---axial-vector---component has
been generated at the one-loop level. Although interesting,
the effect of this term to the decay will be of 
${\cal O}(g_{w}^{4})$ and will be neglected henceforth. 
In this way we arrive at the following form of the 
$\Lambda$-propagator: 
\begin{equation}
     S(p)=\frac
          {A_{0}\gamma^{0}
          -A_{\rm v}{\boldmath\gamma}\cdot{\hat{\bf p}}
          +B}
          {A_{0}^{2}-A_{\rm v}^2-B^{2}} \;.
 \label{LHpropagator}
\end{equation}
where the following quantities have been introduced
\begin{mathletters}
 \begin{eqnarray}
  A_{0}     &\equiv& p^{0\star}-\Sigma_{0}    \;, \\
  A_{\rm v} &\equiv& |{\bf p}|-\Sigma_{\rm v} \;, \\
  B         &\equiv& M_{\Lambda}^{\star}
                              +\Sigma_{\rm s} \;.
 \end{eqnarray}
\end{mathletters}
All these quantities depend (for an on-shell lambda) on the
momentum of the lambda $({\bf p})$ and the density of the
system $(k_{F})$; for simplicity, reference to these
parameters has been suppressed. The medium-modified width 
of the $\Lambda$-hyperon is now extracted by evaluating the
imaginary part of the denominator in Eq.(\ref{LHpropagator})
at the position of the pole. That is,
\begin{equation}
   \Gamma_{\Lambda}=
   -2\,{\cal I}_{m}\left(
   \frac{E_{\Lambda}^{\star}({\bf p})}
	{M_{\Lambda}^{\star}}\Sigma_{0}-
   \frac{|{\bf p}|}{M_{\Lambda}^{\star}}\Sigma_{\rm V}+
   \Sigma_{\rm S}\right)\;.
 \label{LWidth} 
\end{equation}
Since in our model the medium-modified pion propagator depends solely
on the magnitude and not on the direction of ${\bf q}$, the delta
function in Eq.(\ref{ImSigma}) serves to perform the angular 
integrations. This results in the following form for the imaginary
part of the $\Lambda$-self-energy:
\begin{equation}
  {\cal I}_{m}\Sigma_{i}({\bf p})\!=\!
  -\frac{3g_{w}^{2}}{8\pi |{\bf p}|}\!
  \int_{0}^{\infty}\!q\,dq
  \int_{0}^{\infty}\!d\omega\,C_{i}(q,\omega)
  \left[-\frac{1}{\pi}{\cal I}_{m}\,
  \Delta(q,\omega)\right] \;,
 \label{ImSigmas}
\end{equation}
where the $C_{i}$-coefficients are defined by
\begin{equation}
  C_{i}(q,\omega)\!=\!\cases{
   (1\!-\!\kappa^{2})\,M_{N}^{\star}                \;,     
                 & for $i={\rm S}\;;$ \cr
   (1\!+\!\kappa^{2})\,E_{N}^{\star}({\bf p\!-\!q}) \;, 
                 & for $i=0\;;$       \cr
   (1\!+\!\kappa^{2})\,
   \Big(|{\bf p}|-{\bf q}\cdot{\hat{\bf p}}\Big)    \;,  
                 & for $i={\rm V}\;,$ \cr}
 \label{Ccoeffs}
\end{equation}
supplemented with the energy-momentum-conservation relation
\begin{equation}
    E_{N}^{\star}({\bf p\!-\!q})\!=\!
    \sqrt{({\bf p}\!-\!{\bf q})^{2}\!+\!M_{N}^{\star 2}}=
    E_{N}^{(+)}({\bf p\!-\!q})\!-\!V_{N}\!=\!
    E_{\Lambda}^{(+)}({\bf p})\!-\!\omega\!-\!V_{N}\;.
 \label{EMConservation}
\end{equation}
Note that suitable kinematical conditions at the various
interaction vertices impose constrains on the limits 
of integration for both the $q$- and $\omega$-integrals 
in Eq.(\ref{ImSigmas}). For a detailed discussion on these 
limits we refer the reader to the appendix.

\subsection{Pion Propagator II.}
\label{sec:pionpropII}
Equation~(\ref{ImSigmas}) is very general, as it depends solely on
the analytic structure of the pion propagator.  In this section we
discuss in detail the relativistic model used to calculate the pion
propagator---placing special emphasis on the physics underlying its
imaginary part. The imaginary part of the pion self-energy is 
interesting and fundamental as it is related, after a suitable
analysis, to a physical process: the (plane-wave) cross section in 
proton-neutron scattering. There are a variety of physical processes 
that modify the propagation of the pion as it moves through the 
many-body environment. These include the coupling of the pion to 
particle-hole (ph) and nucleon-antinucleon $(N\bar{N})$ excitations. 
In nuclear matter these two kind of excitation have a distinctive
character. For example, pions with spacelike $(q^{2}\!<\!0)$ momenta 
can only ``decay'' into (real) ph-pairs. In contrast, the
$N\bar{N}$-channel opens up at the relatively large timelike momentum
of $q^{2}\!=\!4M_{N}^{\star 2}\!>\!0$.

\begin{figure}
\bigskip
\centerline{
  \psfig{figure=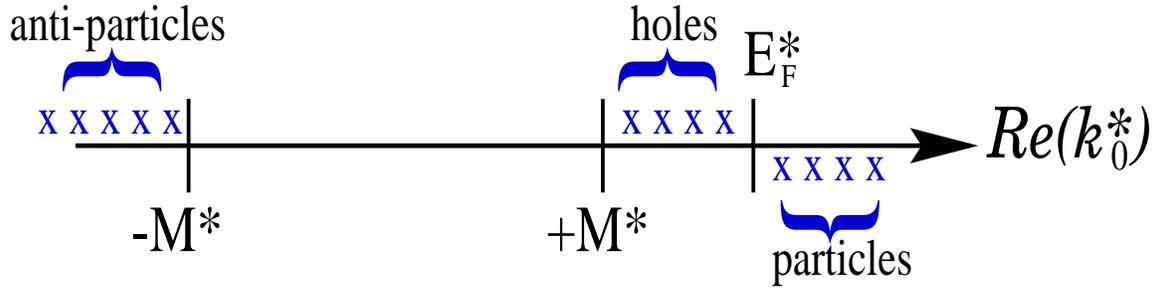,height=1.5in,width=6in,angle=0}}
 \vskip 0.1in
 \caption{Spectral content of the nucleon propagator
          in a relativistic mean-field approximation.}
 \label{fig1}
\end{figure}

\begin{figure}
\bigskip
\centerline{
  \psfig{figure=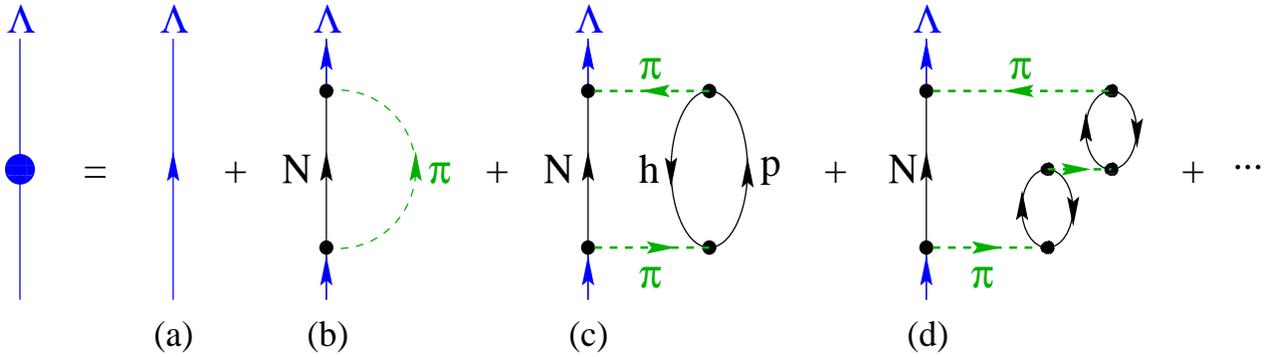,height=1.8in,width=6.6in,angle=0}}
 \vskip 0.1in
 \caption{Feynman diagrams contributing to the 
          in-medium decay width of the $\Lambda$-hyperon.}
 \label{fig2}
\end{figure}

As was discussed in the Introduction, the mesonic decay of the 
lambda---in which an on-shell pion with timelike momentum is 
produced---is strongly Pauli-suppressed in nuclear matter [this 
process is represented by diagram (b) in Fig.~\ref{fig2}]. Recall 
that in free space the nucleon is produced with a momentum of only 
100~MeV, which is well below the Fermi momentum at saturation. 
However, with the closing of the mesonic branch the non-mesonic 
channel opens up. For this nucleon-induced 
($\Lambda N\!\rightarrow\!NN$) decay the exchanged pion is
constrained to have spacelike momentum. Thus, the coupling 
of the pion to spacelike particle-hole excitations becomes of 
paramount importance [this channel is represented by diagram 
(c) in Fig.~\ref{fig2}]. We now compute the medium-modified 
pion propagator in nuclear matter by iterating---to infinite 
order---the lowest-order contribution to the self-energy by 
means of Dyson's equation~\cite{fetwal71}. That is,
\begin{mathletters}
 \begin{eqnarray}
   \Delta(q)&=&\Delta_{0}(q)+\Delta_{0}(q)\Pi_0(q)\Delta(q) 
             = \Delta_{0}(q)+\Delta_{0}(q)\Pi(q)\Delta_{0}(q)\;, 
   \label{PPropagator}  \\
   \Pi(q)&=&\Pi_{0}(q)+\Pi_0(q)\Delta_{0}(q)\Pi(q)
          = \Pi_{0}(q)+\Pi_0(q)\Delta(q)\Pi_{0}(q) \;.
   \label{PSelfenergy}
 \end{eqnarray}
 \label{PPropSelf}
\end{mathletters}
where we have introduced the free pion propagator
\begin{equation}
  \Delta_{0}(q)=\frac{1}{q^{2}-m_{\pi}^{2}+i\eta} \;,
 \label{P0propagator}
\end{equation}
and the lowest-order pion self-energy
\begin{equation} 
 i\Pi_0(q)=\lambda f_{\pi}^{2}
  	\left(\frac{q_{\mu}}{m_{\pi}}\right)
  	\left(\frac{q_{\nu}}{m_{\pi}}\right)
        \int \frac{d^{4}k}{(2\pi)^{4}}{\rm Tr}
	\left[
              \gamma^{\mu}\gamma^{5}G(k+q)
              \gamma^{\nu}\gamma^{5}G(k)
        \right] \;.
 \label{Pselfenergy}
\end{equation}
The factor of $\lambda\!=\!2$ represents the isospin degeneracy 
in symmetric nuclear matter and $f_{\pi}^{2}/4\pi=0.0778$ 
represents the strength of the pseudovector $\pi NN$ vertex.
Further, we have established the equivalence between various ways 
of computing the medium-modified pion propagator. It is the last 
form in Eq.(\ref{PPropagator}) that we use here to write the 
mesonic and non-mesonic contribution to the lambda self-energy:
\begin{mathletters}
 \begin{eqnarray}
   {\cal I}_{m}\Sigma_{i}({\bf p})\Big|_{\rm M} &=&
   -\frac{3g_{w}^{2}}{8\pi |{\bf p}|}\!
   \int_{q_{\rm min}}^{q_{\rm max}}\!q\,dq
   \int_{\omega_{\rm min}}^{\omega_{\rm max}}\!d\omega\,
    C_{i}(q,\omega)
   \left[-\frac{1}{\pi}{\cal I}_{m}\,
   \Delta_{0}(q,\omega)\right] \;, 
  \label{ImSigmasMes} \\
   {\cal I}_{m}\Sigma_{i}({\bf p})\Big|_{\rm N.M.} &=&      
   -\frac{3g_{w}^{2}}{8\pi |{\bf p}|}\!
   \int_{q_{\rm min}}^{q_{\rm max}}\!q\,dq
   \int_{\omega_{\rm min}}^{\omega_{\rm max}}\!d\omega\,
    C_{i}(q,\omega)\,
   \Delta^{2}_{0}(q,\omega)\,S_{L}(q,\omega) \;. 
  \label{ImSigmasNMes}
 \end{eqnarray}
 \label{ImSigmas2}
\end{mathletters}
Note that we have introduced the spin-longitudinal response
$S_{L}(q,\omega)$, to be defined below. The imaginary part 
of the lambda self-energy is obtained by ``cutting'' all 
the Feynman diagrams given in Fig.~\ref{fig1}, or equivalently, 
by putting all intermediate particles on their mass shell. 
All of these diagrams contain a nucleon in the intermediate 
state that must lie above the Fermi surface. Certain diagrams 
[such as diagram (b)] have a pion 
in the intermediate state and contribute to the mesonic
component of the decay. However, as the mesonic component is 
known to be strongly suppressed by the Pauli exclusion principle,
no effort has been made to compute modification to the pion mass 
in the nuclear medium [note that even in the chiral 
($m_{\pi}\!=\!0$) limit the momentum of the outgoing nucleon 
will increase to only 160~MeV; still well below the Fermi momentum
at saturation]. The decay of the lambda into a nucleon and a 
(spacelike) particle-hole excitation [such as in diagram (c)]
constitutes the non-mesonic component of the decay. In this
manner, the non-mesonic component becomes proportional to the 
imaginary part of the lowest-order pion self-energy, or
equivalently, to the spin-longitudinal response:
\begin{equation}
    S^{(0)}_{L}(q,\omega) = - 
    \frac{1}{\pi}{\cal I}_{m}\Pi_{0}(q,\omega) \;.
 \label{spinlong}  
\end{equation}
Finally, diagrams such as in Fig.~\ref{fig1}(d) contribute to 
the non-mesonic decay through the higher-order iteration of the 
lowest-order pion self-energy. These diagrams have been summed 
to all orders using an RPA approach. Note that for the residual
interaction we have included a phenomenological Landau-Migdal
term, as described in detail in Ref.~\cite{horpie93}.

\section{Results}
\label{sec:results}
This section is devoted to the discussion of our results. Before
doing so, however, we clarify the various approximations required 
to extend our nuclear-matter results to finite nuclei. 

The first step in such a procedure is to calculate the ground state of
various closed-shell nuclei (such as ${}^{40}$Ca) in a relativistic
mean-field approximation to the Walecka
model~\cite{walecka74,serwal86}. Once the self-consistent mean fields
are generated, we solve the Dirac equation for a $\Lambda$-hyperon
moving in these potentials; no additional self-consistency is
demanded. Hence, the lambda moves in a mean-field potential that
differs from the corresponding nucleon potential in only two
ways. First, the scalar and vector potentials were scaled by a factor
of 2/3, as the strange quark in the lambda is assumed not to couple to
the sigma nor to the omega meson. Second, by assuming that the
magnetic moment of the lambda is carried exclusively by the strange
quark, a strong tensor mean-field potential is generated. To
illustrate these modifications, we have included in Table~\ref{table1}
a comparison between the nucleon and the lambda single-particle
spectra using a mean-field potential appropriate for ${}^{40}$Ca. For
the case of the $s^{1/2}$ states---which lack spin-orbit
partners---the binding energy for the lambda orbitals is almost 2/3
that of the nucleon.  Yet the most interesting modification is the
near total dilution of spin-orbit effects in the lambda spectrum due
to the strong tensor interaction.

Once the $\Lambda$-spectrum has been obtained, we compute the
effective (or average) vector density sampled by a 
$\Lambda$-hyperon occupying the lowest ($s^{1/2}$) orbital.
That is,
\begin{equation}
 \langle{\rho}_{\rm v}\rangle = 
  \int d^{3}r \rho_{\rm v}(r) 
 \overline{U}_{\Lambda}({\bf r})\gamma^{0}U_{\Lambda}({\bf r})\;.
 \label{rhoavg}
\end{equation}
This is then the nuclear-matter density that is used to compute 
the decay width: 
$\Gamma_{\Lambda}({\bf p};\langle{\rho}_{\rm v}\rangle)$. Finally,
the $\Lambda$-hypernucleus decay width is extracted from the 
convolution of this expression with the momentum distribution
of the lowest orbital:
\begin{equation}
  \Gamma_{\Lambda} = \int \frac{d^{3}p}{(2\pi)^{3}}
  \Gamma_{\Lambda}\Big({\bf p};\langle{\rho}_{\rm v}\rangle\Big)
  \overline{U}_{\Lambda}({\bf p})\gamma^{0}U_{\Lambda}({\bf p})\;.
 \label{FinNucWidth}
\end{equation}

\subsection{Mesonic Decay}
\label{sec:mesdecay}
Computing the mesonic component of the decay is relatively 
straightforward. We start by writing the free pion propagator 
of Eq.(\ref{P0propagator}) in its spectral form 
\begin{equation}
  \Delta_{0}(q)=\frac{1}{2\omega_{\bf q}}
  \left[
   \frac{1}{\omega-\omega_{q}+i\eta}-
   \frac{1}{\omega+\omega_{q}-i\eta}
  \right] \;; \quad
  \omega_{q} \equiv \sqrt{{\bf q}^{2}+m_{\pi}^{2}} \;.
 \label{P0propagator2}
\end{equation}
This yields the following simple form for the mesonic 
component of the self-energy [Eq.(\ref{ImSigmasMes})]
\begin{equation}
   {\cal I}_{m}\Sigma_{i}({\bf p})\Big|_{\rm M} =
   -\frac{3g_{w}^{2}}{16\pi |{\bf p}|}\!
    \int_{\omega_{1}}^{\omega_{2}}
     d{\omega}\,C_{i}(\omega)\;.
 \label{ImSigmasMes2}
\end{equation}
Here the constant $C_{i}$ introduced earlier must be evaluated at 
its ``on-shell'' value $\omega\!\equiv\!\omega_{q}$. Moreover, the 
limits of integration are obtained from simple kinematical
considerations after using the definitions given in 
Eqs.(\ref{wminmax1}) and~(\ref{qminmax}). As the weak $\Lambda N\pi$ 
coupling constants have been chosen to reproduce the width of the 
lambda-hyperon in free space, we obtain in the limit of 
$k_{\rm F}\!=\!0$:
\begin{equation}
  \Gamma_{0}=2.490\!\times\!10^{-12}\,{\rm MeV} \;.
  \label{freewidth}
\end{equation}
Yet the mesonic component of the decay gets modified dramatically
in nuclear matter because of strong Pauli correlations. This can
be seen in Table~\ref{table2} which shows that the mesonic component 
of the decay drops dramatically with baryon number. We will return
to discuss these results in more detail after presenting results
for the non-mesonic component of the decay.

\subsection{Non-Mesonic Decay}
\label{sec:nonmesdecay}
The non-mesonic component of the decay is considerably more
complicated to evaluate. Not only does one need to compute 
the ground-state of uniform nuclear matter but, in addition, 
one must compute its linear (spin-longitudinal) response. 
The linear response of the ground state---contained in the 
imaginary part of the polarization insertion---requires a 
specific representation for the $\pi NN$ vertex. In this 
work we have adopted a pseudovector (PV) representation. 
The pseudovector representation adopted here is just one 
of several possible choices. Indeed, we could have adopted 
instead a pseudoscalar (PS) representation, which is 
guaranteed to be equivalent on-shell. The merit of the PV
representation, however, is that the correct soft-pion limit of
various observables is enforced at the level of the Lagrangian
density, rather than from a sensitive cancellation among various 
Feynman diagrams~\cite{matser82}. 

One of the most interesting results that emerged from our 
earlier work on the pion-nucleon system is the dynamical 
quenching of the $\pi NN$ coupling constant in the nuclear 
medium~\cite{horpie93,dawpie91}. The origin of this quenching 
is intimately related to the the non-conservation of the nucleon 
axial-vector current or, equivalently, to the presence of a
nucleon-mass term. That is,
\begin{equation}
  q_{\mu}\overline{U}({\bf p}')\gamma^{\mu}\gamma^{5}U({\bf p})=
  2M_{N}\overline{U}({\bf p}')\gamma^{5}U({\bf p})\ne0 \;.
 \label{AxialCurrent}
\end{equation}
In the nuclear medium the nucleon mass is reduced---dynamically---from 
its free-space value because of the strong scalar field. Thus,
relative to a ``nonrelativistic'' calculation, by which we mean
that the potentials are small and $M_{N}^{\star}\rightarrow M_{N}$, 
it appears as if the $\pi NN$ coupling constant has been effectively 
reduced in the medium. Moreover, this reduction is density dependent 
and tracks the behavior of the effective nucleon mass; namely, the 
quenching increases with the density of the system. This relativistic 
behavior was responsible for generating no pion condensation---even 
in the absence of short-range correlations~\cite{dawpie91}. More 
importantly, it described correctly\cite{horpie93} the behavior of 
various quasielastic $(\vec{p},\vec{n})$ spin observables measured 
at LAMPF~\cite{McCl92,Chen93}. Measuring a variety of spin observables 
was required for the extraction of the the spin-longitudinal response,
which was instrumental in eliminating pion condensation as a possible
new state of matter. With these results in mind, we now proceed to 
use the same exact relativistic model to compute the non-mesonic 
decay of the $\Lambda$-hyperon.

\begin{figure}
\bigskip
\centerline{
  \psfig{figure=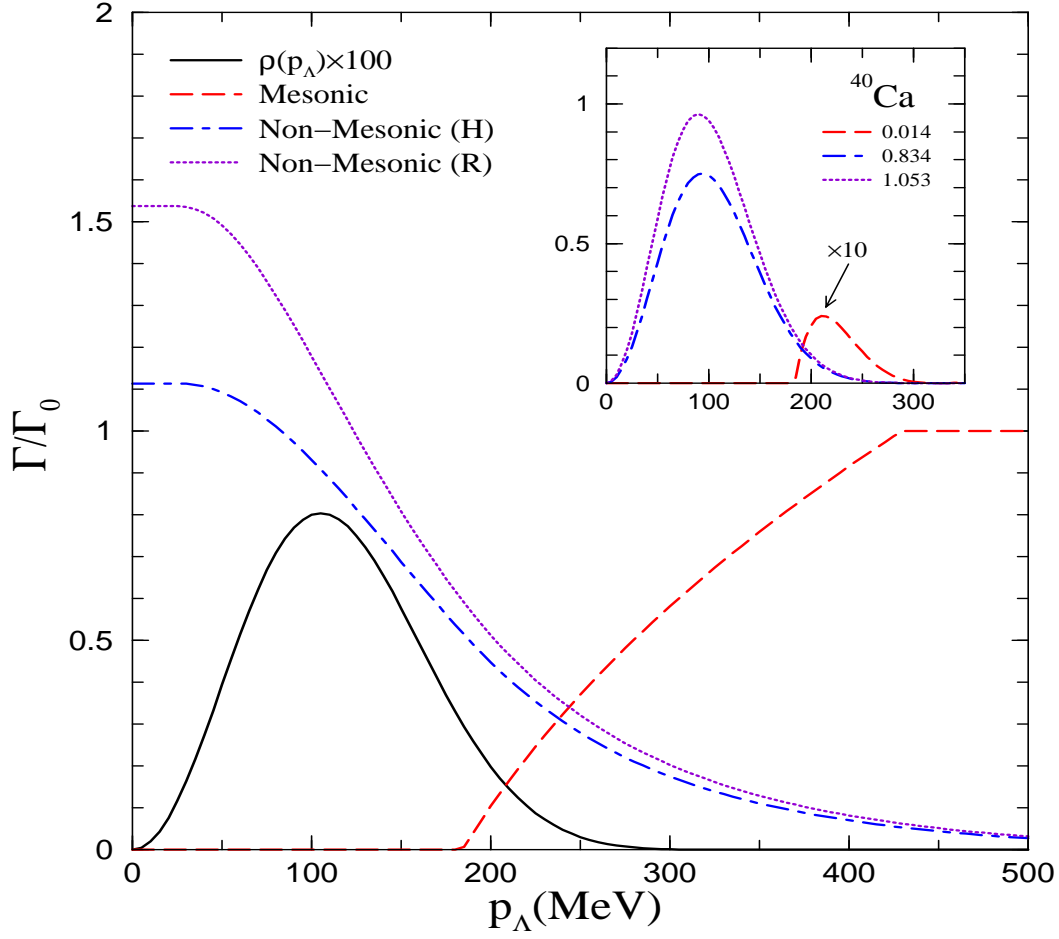,height=5in,width=5.5in,angle=0}}
 \vskip 0.1in
 \caption{Momentum dependence of the $\Lambda$-hyperon decay
          width, normalized to its free-space value. The mesonic 
	  contribution to the width is displayed with the
	  dashed line, while the non-mesonic decay---with and
	  without RPA correlations---is contained in the dotted 
	  and dot-dashed curves, respectively. All calculations
          were done assuming the ``nonrelativistic''
	  ($M_{N}^{\star}\!\rightarrow\! M_{N})$ limit.
	  The inset shows the same curves but now folded with 
	  the momentum distribution of the lowest ($s^{1/2}$) 
	  lambda orbital in ${}^{40}_{\;\Lambda}$Ca.}
 \label{fig3}
\end{figure}

Equation~(\ref{ImSigmasNMes}) suggests that once the spin-longitudinal
response is obtained, computing the different Lorentz components of
the non-mesonic decay becomes straightforward. We present in 
Fig.~\ref{fig3} results for the decay width of the $\Lambda$-hyperon 
at a nuclear-matter density ($k_{\rm F}\!=\!k_{\rm F}^{0}$)
appropriate for ${}^{40}_{\;\Lambda}$Ca (see discussion at the 
beginning of the section). The momentum distribution for the
lowest ($s^{1/2}$) $\Lambda$-orbital is displayed by the solid 
line in the figure; it is normalized so that the area under the 
curve---which has been multiplied by a factor of 100 for 
clarity---is equal to one. The other three curves in the figure 
display the mesonic and non-mesonic components of the decay 
computed in an approach that has fixed all masses to their 
free-space value; this is our best attempt at reproducing
nonrelativistic results. The dashed curve shows the mesonic 
component of the decay as a function of the momentum of the 
lambda. Clearly, the mesonic mode is strongly suppressed 
for momenta below 200 MeV. The curve eventually ``heals'' to its 
free-space value, but too late to make a contribution to the 
in-medium decay; we have computed a mesonic contribution to 
the decay in ${}^{40}$Ca, relative to its free-space value, of 
only 0.014. This number was obtained by folding the decay width 
computed in ($M_{N}^{\star}\!=\!M_{N}$) nuclear matter with the 
momentum distribution of the $s^{1/2}$-orbital (see inset). 
The non-mesonic contributions to the decay, computed without 
and with RPA correlations, are displayed by the dot-dashed and 
dotted lines, respectively. These contributions peak at small 
momenta and, thus, have good overlaps with the momentum distribution. 
The Hartree (or non-RPA) and RPA contributions to the width have 
increased to 0.834 and 1.053, respectively. 

\begin{figure}
\bigskip
\centerline{
  \psfig{figure=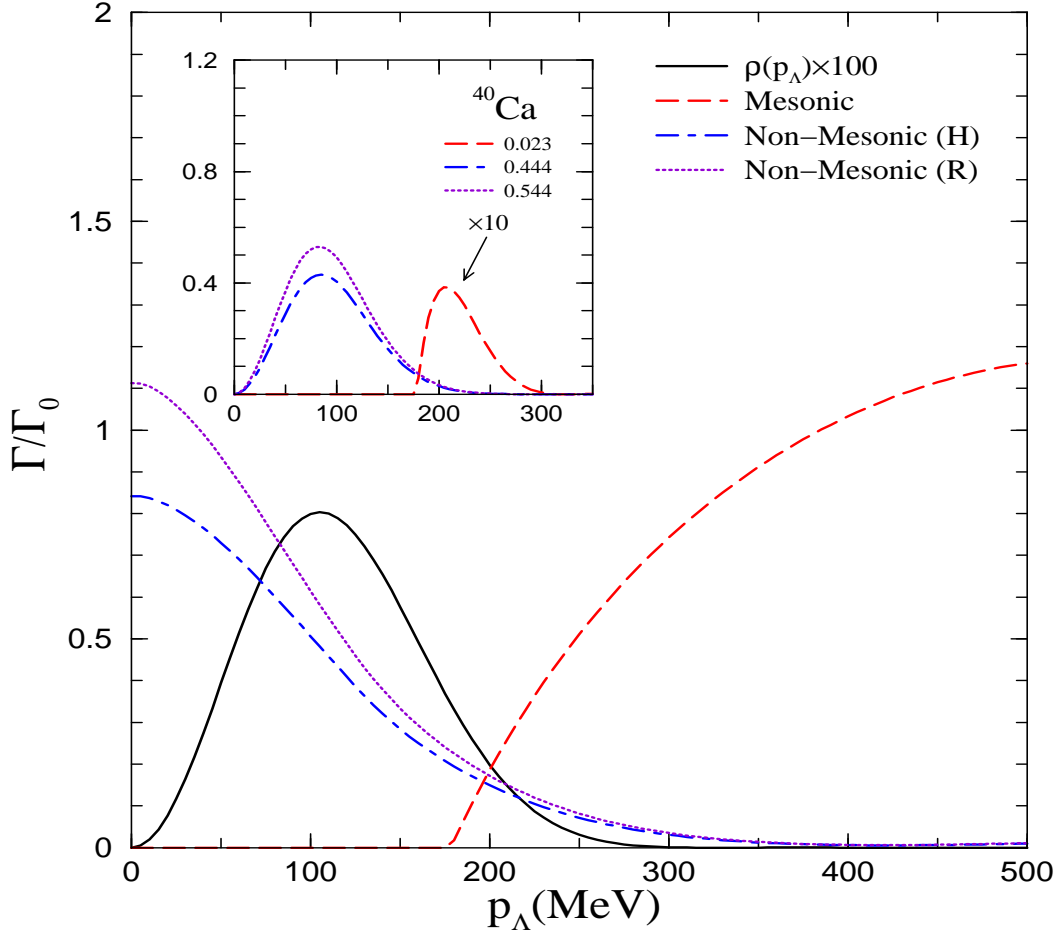,height=5in,width=5.5in,angle=0}}
 \vskip 0.1in
 \caption{Momentum dependence of the $\Lambda$-hyperon decay
          width, normalized to its free-space value. The mesonic 
	  contribution to the width is displayed with the
	  dashed line, while the non-mesonic decay---with and
	  without RPA correlations---is contained in the dotted 
	  and dot-dashed curves, respectively. All calculations
          were done assuming the self-consistent ($M_{N}^{\star}$)
          value for the effective nucleon mass. The inset shows 
	  the same curves but now folded with the momentum 
	  distribution of the lowest ($s^{1/2}$) lambda orbital 
	  in ${}^{40}_{\;\Lambda}$Ca.}
 \label{fig4}
\end{figure}

The same calculation has been repeated in Fig.~\ref{fig4}, but now
with self-consistent nucleon and lambda propagators. That is, 
in-medium masses were used with their values determined 
self-consistently from the mean-field equations~\cite{serwal86}. 
The different scaling of the nucleon and lambda masses with density 
generates a slight change in the mesonic component of the decay, but 
one that it still too small to impact the total decay width. By far 
the most important modification emerges from the dynamical quenching 
of the $\pi NN$ coupling in the nuclear medium~\cite{dawpie91}. 
Indeed, the non-mesonic components of the decay computed in the 
relativistic approach are now reduced to 0.444 and 0.544, respectively.
This large reduction factor, relative to the above 
$M_{N}^{\star}\!=\!M_{N}$ calculation, is consistent with the 
reduction observed in the quasielastic $(\vec{p},\vec{n})$ 
reaction\cite{horpie93}. Moreover, the relativistic RPA result 
obtained here for the total width---$\Gamma/\Gamma_{0}=0.567$---is 
more than a factor of three smaller than the nonrelativistic value
reported by Ramos, Oset, and Salcedo in Ref.~\cite{raossa95}. This 
large reduction factor is not exclusive to ${}^{40}$Ca, but is seen 
in all hypernuclei reported in Table~\ref{table2}. 

It is also relevant to mention that in a very recent publication (of
only a few days old!) a new nonrelativistic evaluation of the
$\Lambda$-decay width has been made~\cite{alber99}. For the case of
${}^{40}$Ca a total decay width (without including the
two-particle---two-hole component of the decay) of
$\Gamma/\Gamma_{0}\!=\!1.08$ has been reported. This represents a
reduction of about 35-40\% relative to the value presented in
Ref.~\cite{raossa95}. Yet, it is essentially identical to our
``nonrelativistic'' ($M_{N}^{\star}\!=\!M_{N}$) value of
$\Gamma/\Gamma_{0}\!=\!1.07$. However, had we used the same
$M_{N}^{\star}\!=\!M_{N}$ model to study the quasielastic
$(\vec{p},\vec{n})$ observables~\cite{McCl92,Chen93}, we would have
grossly overestimated the observed spin-longitudinal
response~\cite{horpie93}.

\section{Conclusions}
\label{sec:concl}
We have computed for the first time the decay width of
$\Lambda$-hypernuclei in a relativistic mean-field approximation to
the Walecka model. Most of the formalism presented here is quite
general, as it relies only on the analytic structure of various
propagators. The calculation was performed in two stages. First,
ground-state properties of various closed-shell nuclei were computed
self-consistently. With the mean-fields potentials in hand, the Dirac
equation for a $\Lambda$-hyperon was solved with scaled scalar and
vector potentials. A scaling factor of 2/3 was introduced from the
assumption that the strange quark in the lambda does not to couple to
the sigma nor to the omega meson. Moreover, by assuming that the
magnetic moment of the $\Lambda$-hyperon is carried solely by the
strange quark, a tensor mean-field potential was introduced that
diluted---almost completely---all spin-orbit effects.  Second, the
mesonic and non-mesonic components of the decay width---with and
without RPA correlations---were computed in infinite nuclear
matter. The uniform nuclear-matter density was estimated from
evaluating the average density sampled by a $\Lambda$-hyperon
occupying the lowest ($s^{1/2}$) orbital. Finally, the 
nuclear-matter results were folded with the momentum distribution 
of the lowest $\Lambda$-orbital to compute the finite-hypernuclear 
width.

As expected, the mesonic component of the decay was strongly
suppressed by Pauli correlations. Indeed, for the values of the
nuclear-matter density adopted here, the decay of $\Lambda$-hyperons
with momenta less than 200 MeV was strictly forbidden. Since in the
Walecka mean-field model the momentum components of the
$\Lambda$-hyperon beyond 200 MeV are small, the strong suppression of
the mesonic mode ensued. In contrast, the non-mesonic component of the
decay is not Pauli-blocked. In the non-mesonic mode the
$\Lambda$-hyperon decays into a nucleon and a pion-like pair with
spacelike momenta. Information on these kind of pion-like excitations
is contained in the nuclear spin-longitudinal response. This
fundamental nuclear response is accessible through charge-exchange
$(\vec{p},\vec{n})$ reactions.  Indeed, high-precision measurements
made at the NTOF facility at LAMPF were instrumental in answering
some fundamental nuclear-physics questions~\cite{McCl92,Chen93}. 
For example, these experiments showed conclusively, by extracting
the spin-longitudinal response, that the long-sought pion-condensed
state does not exist. In our relativistic model the absence of pion
condensation was attributed to a dynamical quenching of the $\pi NN$
coupling constant in the nuclear medium~\cite{horpie93,dawpie91}. 

In the present work we have computed the non-mesonic component of the
decay---using the same relativistic RPA model that was used to confirm
the LAMPF measurements. The dynamical quenching of the $\pi NN$
coupling constant in the medium was now responsible for generating
decay widths for $\Lambda$-hypernuclei that were two to three times
smaller than those obtained by existing nonrelativistic
calculations~\cite{raossa95,alber99}---and considerably smaller than
measured experimentally~\cite{szyman91}. This suggests that other
modes of excitation, besides pion-like modes, might be important in
understanding the decay width of
$\Lambda$-hypernuclei~\cite{dubach86,dubach96,ramben95}.

The study of such modes seems both interesting and necessary. For
example, a weak $\Lambda N\omega$ vertex will induce a coupling to
isoscalar-vector modes. As in the pion case, this additional
contribution to the non-mesonic width can be related to a well-known
nuclear response: the longitudinal response measured in electron
scattering. Hence, its impact on the decay will be also strongly
constrained. Searching for the underlying dynamics behind the decay
width of $\Lambda$-hypernuclei---while at the same time maintaining
consistency with the wealth of existing nuclear-response data---will
be the subject of a future report.

\acknowledgments
This work was supported by the DOE under
Contracts Nos. DE-FC05-85ER250000, DE-FG05-92ER40750.

\vfill\eject

\appendix
\section{Kinematical Constraints at the Vertices}
\label{sec:appendix} 
The limits of integration in the $q$- and $\omega$-integrals of
Eq.(\ref{ImSigmas2}) are constrained by energy-momentum 
conservation at the various interacting vertices. We now examine 
the kinematical constraints imposed at each vertex separately.

\subsection{Kinematical Constraints at the 
             Lambda-Nucleon-Pion Vertex}
\label{sec:appendix1} 
Energy-momentum conservation at the $\Lambda N\pi$ vertex demands
that an on-shell lambda-hyperon ``decays'' into an on-shell 
nucleon---above the Fermi surface---and, in principle, an off-shell
pion. That is,
\begin{equation}
  \omega = E_{\Lambda}^{(+)}({\bf p})-
           E_{N}^{(+)}({\bf p}\!-\!{\bf q}) \ge 0 \;.
 \label{vertex1}
\end{equation}
Moreover, as the angular integration over ${\bf q}$ can always
be performed, the following limits of integration ensue:
\begin{mathletters}
 \begin{eqnarray}
   \omega_{\rm min}&=&E_{\Lambda}^{(+)}\!-\!
   \max\left(E_{N}^{(+)}(k_{F}),E_{N}^{(+)}(p+q)\right)\;, \\
   \omega_{\rm max}&=&E_{\Lambda}^{(+)}\!-\!
   \max\left(E_{N}^{(+)}(k_{F}),E_{N}^{(+)}(p-q)\right)\;,
 \end{eqnarray}
 \label{wminmax1}
\end{mathletters}
where $E_{N}^{(+)}(p\pm q)=\sqrt{(|{\bf p}|\pm q)^{2}\!+\!
M_{N}^{\star 2}}+V_{N}$. Note that the above relations
impose constraints on the $q$-integral as well. For example,
the $\omega$-integral will vanish unless
$\omega_{\rm max}\!>\!\omega_{\rm min}$. Further,
$\omega_{\rm min}$ is constrained to be greater than
or equal to zero. Thus, from these two conditions the
following limits are obtained:  
\begin{equation}
   q_{\rm min}=\max\left(0,k_{F}\!-\!|{\bf p}|\right) \;, \quad
   q_{\rm max}=\max\left(0,\sqrt{(E_{\Lambda}^{(+)}\!-\!V_{N})^{2}
              -M_{N}^{\star 2}}-|{\bf p}|\right) \;.
 \label{qminmax}
\end{equation}

\subsection{Kinematical Constraints at the 
	    Particle-Hole-Pion Vertex}
\label{sec:appendix2} 
In the case of the non-mesonic decay of the lambda-hyperon, 
additional constraints---at least in nuclear matter---follow 
from enforcing energy-momentum conservation at the ph$\pi$ vertex.
Indeed, the energy of the exchanged (spacelike) pion is given by:
\begin{equation}
  \omega = E_{N}^{(+)}({\bf k}+{\bf q})
         - E_{N}^{(+)}({\bf k}) \ge 0 \;,
 \label{vertex2}
\end{equation}
where $E_{N}^{(+)}({\bf k}+{\bf q})\!>\!E_{N}^{(+)}(k_{F})$ 
denotes the energy of the particle and 
$E_{N}^{(+)}({\bf k})\!<\!E_{N}^{(+)}(k_{F})$ is the energy
of the hole. As in the above case, the most stringent 
constraints emerge in the case in which the particle, the 
hole, and the pion are collinear. We obtain,
\begin{mathletters} 
 \begin{eqnarray}
   \omega_{\rm min}&=&\max\left(0,
         \sqrt{(|{\bf q}|\!-\!k_{F})^{2}+M_{N}^{\star 2}}
        -\sqrt{k_{F}^{2}+M_{N}^{\star 2}}\right) \;, \\
   \omega_{\rm max}&=&
         \sqrt{(|{\bf q}|\!+\!k_{F})^{2}+M_{N}^{\star 2}}
        -\sqrt{k_{F}^{2}+M_{N}^{\star 2}} \;.
 \end{eqnarray}
 \label{wminmax2}
\end{mathletters}
The resulting limits of integration for the $\omega$-integral, 
then follow from enforcing Eqs.(\ref{wminmax1}) 
and~(\ref{wminmax2}) simultaneously.


%
%
\vfill\eject
\vfill\eject
\vfill\eject
%
%
%
%
\mediumtext
 \begin{table}
  \caption{Single-particle binding energies (in MeV)
	   for a nucleon or a lambda in the 
	   self-consistent mean-field of ${}^{40}$Ca.
           Quark-model estimates were used for 
           $g_{\sigma\Lambda\Lambda}$ and 
           $g_{\omega\Lambda\Lambda}$. See text for 
	   details.} 
   \begin{tabular}{cccc}
     Orbital   &     N    &  $\Lambda$  \\
   \tableline
    $s^{1/2}$ & -55.390  & -36.274  \\
   \tableline
    $p^{3/2}$ & -38.903  & -24.030  \\
    $p^{1/2}$ & -33.181  & -23.591  \\
   \tableline
    $d^{5/2}$ & -22.751  & -11.769  \\
    $s^{1/2}$ & -14.389  & - 9.838  \\
    $d^{3/2}$ & -13.876  & -11.099  \\
   \end{tabular}
  \label{table1}
 \end{table}
\mediumtext
 \begin{table}
  \caption{Hypernuclear decay widths in a relativistic mean-field
           approximation to the Walecka model. The third column
           contains the mesonic contribution to the decay width,
           while the fourth and fifth columns display the 
	   dominant non-mesonic and total decay widths, 
	   respectively. The last two columns contain the 
           nonrelativistic results, without including the 
	   two-particle---two-hole component of the decay, 
	   as computed in Ref.~{\protect\cite{raossa95}} and 
	   in Ref.~{\protect\cite{alber99}}, respectively.}
   \begin{tabular}{ccccccc}
    Nucleus & $\langle k_{\rm F}\rangle\!/\!k_{\rm F}^{0}$ 
    & $\Gamma_{\rm M}/\Gamma_{0}$
    & $\Gamma_{\rm N.M.} /\Gamma_{0}$   
    & $\Gamma_{\rm Total}/\Gamma_{0}$  
    & $\Gamma/\Gamma_{0}$~\cite{raossa95} 
    & $\Gamma/\Gamma_{0}$~\cite{alber99} \\
   \tableline
    ${}^{12}_{\;\Lambda}$C  & 0.991 & 0.112 & 0.413 
                            & 0.525 & 1.76  & 1.07  \\
    ${}^{16}_{\;\Lambda}$O  & 0.956 & 0.099 & 0.497  
                            & 0.596 & 1.78  & ---   \\
    ${}^{28}_{\;\Lambda}$Si & 1.013 & 0.053 & 0.479 
                            & 0.532 & ---   & 1.09  \\
    ${}^{32}_{\;\Lambda}$S  & 1.024 & 0.048 & 0.491 
                            & 0.539 & ---   & ---   \\
    ${}^{40}_{\;\Lambda}$Ca & 1.000 & 0.023 & 0.544 
                            & 0.567 & 1.79  & 1.08  \\
   \end{tabular}
  \label{table2}
 \end{table}

\end{document}